\documentstyle[amssymb,rmp,aps]{revtex}

\begin{document}
\title{The Rotational Structure of Molecules In the Quantum $^{4}$He and $^{3}$He
Liquids}
\author{V. S. Babichenko and Yu. Kagan }
\address{RSC ''Kurchatov Institute'', Moscow 123182, Russia. \\
e-mail: babichen@kurm.polyn.kiae.su}
\date{}
\maketitle

\begin{abstract}
It is shown that the drastic distinction in the rotational structure of
molecules in the $^{4}$He and $^{3}$He liquids observed in [1] is due to the
difference in the spectral density of excitations regardless of the
hydrodynamic properties. The large width of rotational levels in the $^{3}$%
He is determined by particle-hole excitations whereas the small density of
phonon excitations results only in the small broadening of levels in the $%
^{4}$He.
\end{abstract}

\bigskip

1. In a recent paper $\left[ 1\right] $ \ a well resolved rotational
structure \ has been observed in the infra-red spectrum of the linear OCS
molecule embedded in a $^{4}$He cluster at T%
\mbox{$<$}%
T$_{\lambda }.$ In fact, this group has observed this phenomenon earlier on
the example of the SF$_{6}$ molecule $\left[ 2\right] $ \ but with less
resolution. However, the most impressive and interesting aspect is the
observation in the same work $\left[ 1\right] $ \ that the rotational
structure is completely smeared out in pure $^{3}$He clusters even at lower
temperatures. It would be natural to connect these results simply with
superfluidity in the $^{4}$He and its lack in the $^{3}$He at the
temperature of measurements. However, the hydrodynamic aspect of
superfluidity with the corresponding macroscopic kinetic coefficients is not
applicable to this problem. This is connected with the fact that the mean
free path of excitations in the quantum Bose- and Fermi- liquids under those
conditions is large compared with the size of a molecule. However, it is the
coupling with the excitations that determines the dissipative picture of the
rotational spectrum. The coupling with the ground state of the liquid
results in a renormalization of the molecular moment of inertia which is
observed in the experiment.

The character and spectral properties of excitations are the specific
features of a quantum liquid at low temperatures. The Bose liquid exhibits
phonon excitations whereas the Fermi liquid has mainly pair particle-hole
excitations. In the following we will show that this distinction in the
spectral properties of the excitation branches determines the difference in
the width of the rotational levels in these liquids by several orders of the
magnitude. This explains the different spectra observed in $\left[ 1\right] $
\ for these two quantum liquids which differ only in the statistics and have
practically the same coupling with the molecule. The connection with
superfluidity is due to the fact that the statistics plays the decisive role
in the appearance of this unique property.

2. To reveal the principle features of the interaction with the quantum
liquids, the molecule is assumed to be a rigid rotator. The Hamiltonian of
the interaction in this case can be written in the following form

\begin{equation}
H_{int}=\int d^{3}\overrightarrow{r}U\left( \overrightarrow{r}-%
\overrightarrow{R}\right) \widehat{\rho }\left( \overrightarrow{r}\right)
=\sum\limits_{\overrightarrow{k}}U\left( \overrightarrow{k}\right) \exp
\left( -i\overrightarrow{k}\overrightarrow{R}\right) \widehat{\rho }\left( 
\overrightarrow{k}\right)
\end{equation}
Here $\widehat{\rho }\left( \overrightarrow{r}\right) $ is the operator of
the liquid density, $U\left( \overrightarrow{r}-\overrightarrow{R}\right) $
is the potential for the coupling of the rotator with the liquid , and $%
\overrightarrow{R}$ is the coordinate of the rotator 
\[
\overrightarrow{R}=R_{0}\overrightarrow{n}, 
\]
where $R_{0}=const$, and $\overrightarrow{n}$ is the spatial unit vector .

First we suppose that the volume of the liquid is very large. Later on we
will consider the role of the finite size of the system.

By assuming the interaction (1) to be weak it is possible to use
perturbation theory. The probability of the transition between the
rotational levels j$_{0}$ and j$_{1}$ with simultanious excitation of the
liquid is then given by

\begin{equation}
W_{j_{0}j_{1}}=\frac{2\pi }{\hbar }\int \frac{d^{3}k}{\left( 2\pi \right)
^{3}}\left| U\left( \overrightarrow{k}\right) \right| ^{2}S\left( 
\overrightarrow{k},\varepsilon _{j_{0}j_{1}}\right)
\sum\limits_{m_{1}}\left| \left( \exp \left( -i\overrightarrow{k}%
\overrightarrow{R}\right) \right) _{j_{0},m_{0};j_{1},m_{1}}\right| ^{2}
\end{equation}
where $S\left( \overrightarrow{k},\varepsilon _{j_{0}j_{1}}\right) $ is the
dynamic form factor which only depends on the properties of the liquid and
is given by

\begin{equation}
S\left( \overrightarrow{k},\varepsilon _{j_{0}j_{1}}\right) =\frac{1}{Z}%
\sum\limits_{s,s_{1}}\exp \left( -\frac{E_{s}}{T}\right) \left| \left( 
\widehat{\rho }\left( \overrightarrow{k}\right) \right) _{s,s_{1}}\right|
^{2}\delta \left( \varepsilon _{j_{0}j_{1}}-E_{s_{1}}+E_{s}\right)
\end{equation}
Here $s$ and $s_{1}$ denote the states of the liquid which correspond to the
energies $E_{s}$ and $E_{s_{1}},$ respectively, and 
\begin{equation}
\varepsilon _{j_{0}j_{1}}=\varepsilon _{j_{0}}-\varepsilon _{j_{1}}.
\end{equation}
where $\varepsilon _{j}$ is the energy of the rotator level with the moment $%
j$, \ $Z\ $ is the partition function. In Eqs. (2) and (3) it is implicitly
assumed that the liquid is homogeneous.

3. Let us consider a Bose liquid supposing that $\mid \varepsilon
_{j_{0}j_{1}}\mid ,T<<T_{\lambda }$. To determine the function S, we employ
the known expression connecting density operator of Bose liquid with the
creation $\widehat{b}_{\overrightarrow{k}}^{+}$ and annihilation $\widehat{b}%
_{\overrightarrow{k}}$ phonon operators $\left[ 3\right] $ . 
\begin{equation}
\widehat{\rho }\left( \overrightarrow{r}\right) =i\sum\limits_{%
\overrightarrow{q}}\left( \frac{\hbar qn}{2m_{0}u}\right) ^{1/2}\left( 
\widehat{b}_{\overrightarrow{q}}\exp \left( i\overrightarrow{q}%
\overrightarrow{r}\right) -\widehat{b}_{\overrightarrow{q}}^{+}\exp \left( -i%
\overrightarrow{q}\overrightarrow{r}\right) \right)
\end{equation}
where $n$ is the density of a liquid, $u$ is the velocity of sound and $%
m_{0} $ is the bare mass of the atoms of a Bose liquid. \ (The volume $V$=1)

After some calculations we find 
\begin{equation}
S\left( \overrightarrow{k},\varepsilon _{j_{0}j_{1}}\right) =\frac{\mid
\varepsilon _{j_{0}j_{1}}\mid n}{2m_{0}u^{2}}B\left( \varepsilon
_{j_{0}j_{1}};T\right) \delta \left( \mid \varepsilon _{j_{0}j_{1}}\mid
-\hbar ku\right)
\end{equation}
where

\[
B\left( \varepsilon _{j_{0}j_{1}};T\right) =\left( n_{B}\left( \varepsilon
_{j_{0}j_{1}}\right) +1\right) \theta \left( \varepsilon
_{j_{0}j_{1}}\right) +n_{B}\left( -\varepsilon _{j_{0}j_{1}}\right) \theta
\left( -\varepsilon _{j_{0}j_{1}}\right) 
\]
and

\[
n_{B}\left( \varepsilon \right) =\frac{1}{\exp \left( \frac{\varepsilon }{T}%
\right) -1} 
\]

Under the conditions of the experiment $kR_{0}=\frac{\mid \varepsilon
_{j_{0}j_{1}}\mid }{\hbar u}R_{0}<<1.$ This allows us to replace $U\left( 
\overrightarrow{k}\right) $ by $U\left( 0\right) \equiv U_{0}$ in Eq. (2)
and to retain only the linear term in $\overrightarrow{R}$ for the expansion
of exponent. The direct calculation yields

\begin{equation}
\sum\limits_{m_{1}}\left| \left( \overrightarrow{k}\overrightarrow{R}\right)
_{j_{0},m_{0};j_{1},m_{1}}\right| ^{2}=k^{2}R_{0}^{2}f_{j_{0}j_{1}}
\end{equation}

where the expression for $f_{j_{0}j_{1}}$ has the form

\begin{equation}
f_{j_{0}j_{1}}=\frac{j_{0}^{2}}{\left( 2j_{0}-1\right) \left(
2j_{0}+1\right) }\text{ \ }\left( \text{if \ }j_{1}=j_{0}\right) ,
\end{equation}

\[
f_{j_{0}j_{1}}=\frac{\left( j_{0}+1\right) ^{2}}{\left( 2j_{0}+1\right)
\left( 2j_{0}+3\right) }\text{ \ \ }\left( \text{if \ }j_{1}=j_{0}+1\right)
. 
\]

For the $j_{0}=1\rightarrow j_{1}=0$ transition, the coefficient $f_{01}$
equals $1/3$. Substituting Eqs. (6) and (7) into Eq. (2), we obtain for the
transition probability

\begin{equation}
W_{j_{0}j_{1}}^{B}=\left| U_{0}\right| ^{2}\frac{n\left| \varepsilon
_{j_{0}j_{1}}\right| ^{5}R_{0}^{2}}{2\pi m_{0}u^{7}\hbar ^{6}}%
f_{j_{0}j_{1}}B\left( \varepsilon _{j_{0}j_{1}};T\right) .
\end{equation}

4. Let us turn now to the case of the Fermi liquid. We are interested only
in the conditions satisfying the inequalities

\begin{equation}
\mid \varepsilon _{j_{0}j_{1}}\mid ,T<<\varepsilon _{F}
\end{equation}

In this case we will employ the Landau theory of the Fermi liquid $\left[ 3%
\right] ,$ $\left[ 4\right] $ for the estimate of the transition
probability. Within the framework of this theory under the condition Eq.
(10) \ and $1/\tau <<kV_{F}$ , $\tau $ being the typical collision time of
excitations, the dynamical form factor Eq. (3) can be written as (see $\left[
4\right] $ ) 
\begin{equation}
S\left( \overrightarrow{k},\varepsilon _{j_{0}j_{1}}\right) =\frac{\left|
\varepsilon _{j_{0}j_{1}}\right| }{2\hbar kv_{F}}\frac{\nu _{F}}{\left(
1+F_{0}^{s}\right) ^{2}}B\left( \varepsilon _{j_{0}j_{1}};T\right) \theta
\left( 2k_{F}-k\right) ,
\end{equation}
where the density of states at the Fermi surface $\nu _{F}$ is given by 
\begin{equation}
\nu _{F}=\frac{m^{\ast }p_{F}}{\pi ^{2}\hbar ^{3}}=\frac{m_{0}p_{F}}{\pi
^{2}\hbar ^{3}}\left( 1+\frac{F_{1}^{s}}{3}\right)
\end{equation}
Here $F_{0}^{s}$ and $F_{1}^{s}$ are the dimensionless parameters of the
Fermi liquid theory and characterize the interaction between quasiparticles.

In the limit $k_{F}R_{0}<<1$ we can use the relations Eq. (7) and Eq. (8).
Substituting Eqs. (12) and (7) into Eq. (2) we find

\begin{equation}
W_{j_{0}j_{1}}^{F}=\left| U_{0}\right| ^{2}\frac{2\left| \varepsilon
_{j_{0}j_{1}}\right| R_{0}^{2}}{\pi ^{3}\hbar ^{5}}\frac{m^{\ast 2}}{\left(
1+F_{0}^{s}\right) ^{2}}R_{0}^{2}k_{F}^{4}f_{j_{0}j_{1}}B\left( \varepsilon
_{j_{0}j_{1}};T\right)
\end{equation}

In the general case of an arbitrary value of $k_{F}R_{0}$ one can employ the
expansion 
\[
\exp \left( i\overrightarrow{k}\overrightarrow{R}\right) =4\pi
\sum\limits_{l=0}\sum\limits_{m=-l}^{l}i^{l}\sqrt{\frac{\pi }{2kR_{0}}}%
J_{l+1/2}\left( kR_{0}\right) Y_{lm}\left( \overrightarrow{n}\right)
Y_{lm}\left( \frac{\overrightarrow{k}}{k}\right) 
\]
where $J_{l+1/2}$ are the Bessel functions and $Y_{lm}$ are the standard
spherical functions.

The dynamical form factor $S$ in Eq. (11) depends only on the modulus $%
\overrightarrow{k}$ . That is why the integration over $d\Omega _{%
\overrightarrow{k}}$ conserves only the terms diagonal in $l,m$ in Eq. (2).
\ The calculation of the matrix elements from $Y_{lm}\left( \overrightarrow{n%
}\right) $ is performed in the standard manner $\left[ 5\right] $. As a
result, with the substitution of $U\left( \overrightarrow{k}\right) $ for $%
U_{0}$ we find for the transition probability

\begin{eqnarray}
&&W_{j_{0}j_{1}}^{F}=\left| U_{0}\right| ^{2}\frac{\left| \varepsilon
_{j_{0}j_{1}}\right| m^{\ast 2}}{2\pi ^{3}\hbar ^{5}\left(
1+F_{0}^{s}\right) ^{2}}B\left( \varepsilon _{j_{0}j_{1}};T\right)
\sum\limits_{l,m}\left( 2j_{0}+1\right) \left( 2j_{1}+1\right) \left(
2l+1\right) \times  \nonumber \\
&&\left( 
\begin{array}{ccc}
j_{1} & l & j_{0} \\ 
-m & m-m_{0} & m_{0}
\end{array}
\right) ^{2}\left( 
\begin{array}{ccc}
j_{1} & l & j_{0} \\ 
0 & 0 & 0
\end{array}
\right) ^{2}\int\limits_{0}^{2k_{F}}dkk\left( \sqrt{\frac{\pi }{2kR_{0}}}%
J_{l+1/2}\left( kR_{0}\right) \right) ^{2}
\end{eqnarray}
where the symbols $\left( 
\begin{array}{ccc}
j_{1} & l & j_{0} \\ 
-m & m-m_{0} & m_{0}
\end{array}
\right) $ are the standard 3j-symbols $\left[ 5\right] $.

In the opposite limit $k_{F}R_{0}>>1$ we have

\[
\int\limits_{0}^{2k_{F}}dkk\left( \sqrt{\frac{\pi }{2kR_{0}}}J_{l+1/2}\left(
kR_{0}\right) \right) ^{2}=\frac{1}{2R_{0}^{2}}\ln \left( 2k_{F}R_{0}\right) 
\]

To simplify the comparison of the results in the cases of Bose and Fermi
liquids, we consider the transition between the levels of $j_{0}=1$ and $%
j_{1}=0$. In this case Eq. (14) can be simplified and in the $k_{F}R_{0}>>1$
limit is given by

\begin{equation}
W_{10}^{F}=\left| U_{0}\right| ^{2}\frac{\left| \varepsilon
_{j_{0}j_{1}}\right| m^{\ast 2}}{4\pi ^{3}\hbar ^{5}\left(
1+F_{0}^{s}\right) ^{2}R_{0}^{2}}\ln \left( 2p_{F}R_{0}\right) B\left(
\varepsilon _{01};T\right)
\end{equation}

5. In order to display the distinctions in the dissipative broadening of the
rotational spectrum lines , it is sufficient to compare the probability of
transitions between levels with $j_{0}=1$ and $j_{1}=0$ in the both cases.
In this case, according to (8), we have $f_{10}=1/3$. Since the inequality $%
k_{F}R_{0}>1$ is more typical, we use Eq. (15) for the Fermi liquid. (For $%
k_{F}R_{0}\thicksim 1,$ Eqs. (15) and (13) give \ similar values).
Determining the ratio of Eq. (9) to Eq. (15), we have

\begin{equation}
\frac{W_{10}^{B}}{W_{10}^{F}}\thickapprox \frac{2\pi ^{2}}{3}\left(
1+F_{0}^{s}\right) ^{2}\left( nR_{0}^{3}\right) \left( \frac{\varepsilon
_{01}}{mu^{2}}\right) ^{2}\left( \frac{m_{0}}{m^{\ast }}\right) \left( \frac{%
R_{0}m_{0}u}{\hbar \ln \left( 2k_{F}R_{0}\right) }\right)
\end{equation}

Taking $^{3}$He and $^{4}$He at zero pressure, one can use the following
experimental values of the parameters in Eq. (16) (see, e.g., $\left[ 6%
\right] $, $\left[ 5\right] $ ) for $^{3}$He : $F_{0}^{s}=9.3,$ \ $\frac{%
m^{\ast }}{m_{0}}=2.8,$ \ \ $k_{F}=0.8\times 10^{8}cm^{-1},$ and for $^{4}$%
He : $n=2.3\times 10^{22}1/cm^{3},$ \ $u=2.4\times 10cm/s,$ \ $m_{0}\equiv
m_{4}$ . For the OCS molecule $\left[ 1\right] $ \ the difference in the
rotational energies is $\varepsilon _{01}\thicksim 0.15cm^{-1},$ the
parameter $R_{0}$ is $R_{0}\thicksim 4\div 5$\AA\ and with these parametres 
\begin{equation}
\frac{W_{10}^{B}}{W_{10}^{F}}\thicksim 10^{-5}
\end{equation}

Naturally, this is an approximate result. However the ratio is so small that
even a more precise calculation cannot change the result that the linewidth
in $^{3}$He greatly exceeds \ the magnitude of the linewidth in $^{4}$He.

6. In connection with these results several remarks are in order. For the
Fermi liquid only the dissipative contribution due to particle-hole
excitations was taken into account. The Bose excitations, i.e., usual and
zero sound, are strongly damped at the intermediate temperatures $T\thicksim
\varepsilon _{0}\thicksim 0.1K$ of the experiment. But even disregarding
this fact, the small phase volume characteristic for the Bose excitations
predetermines the negligible role of the Bose excitations in the Fermi
liquid (just due to the same reason the effect in $^{4}$He is small). Note
that this contribution would only further diminish only the ratio Eq. (16).

For the analysis of the transition in the Fermi liquid, we use the classical
variant of the Landau theory which assumes that the parameters $F_{0}^{s}$
and $F_{1}^{s}$ ( see Eq. (11) and (12) ) are independent of $k$. The modern
theories of the Fermi liquid analyze the effective dependence of these
parameters on the momentum $\overrightarrow{k}$ $\left[ 6\right] $, $\left[ 7%
\right] $ . However the fact that the parameters $F_{0}^{s}$ and $F_{1}^{s}$
decrease within the interval between $k_{F}$ and $2k_{F}$ does not change
noticeably the estimate (17). Certain growth of $F_{0}^{s}$ with the fall of 
$F_{1}^{s}$ within the $0\div k_{F}$ interval, predicted in $\left[ 7\right] 
$, can result in the limited enhancement of the estimate Eq. (17). Another
small increase of the ratio Eq. (16) may be associated with the fact that $%
U\left( 2k_{F}\right) <U\left( 0\right) .$

The renormalization of the moment of inertia J of a molecule in the quantum
liquid is connected in decisive degree with the adiabatic reversible
coupling . The renormalization of J should be slightly different in the two
liquids , although the difference in densities should manifest itself. For
the calculation of the ratio in Eq. (16), we used the experimental value of $%
\varepsilon _{01}$ which enters strongly into the value $W_{10}^{B}\thicksim
\left( \varepsilon _{0}\right) ^{5}$ (Eq.(9)) in $^{4}$He. \ 

Moreover in Eq. (16) identical temperatures were assumed for both liquids.
In the $^{3}$He experiments $\left[ 1\right] $ the temperature is about $%
0.15K,$ \ $T\thicksim 0.37K$ in the $^{4}$He . Taking into account this fact
results in the increase of the ratio Eq. (16) is a factor of about two due
to the term $1+n_{B}$.

To summary, it is easy to understand that, though the genuine ratio (16) may
increase compared with the estimate (17), this ratio remains very small.

All the above results are derived assuming a weak interaction. In fact, this
means only that the width of the rotational levels should be small compared
with the spacing between the levels. The experimental width of the
nonresolved peak in $^{3}$He is about 0.1cm$^{-1}$, in other words, of the
order of $\varepsilon _{0}$. This means that only a qualitative comparison
with the theoretical estimate is possible. From the experimental point of
view it would be to have measurements at various temperatures also above $%
T_{\lambda }$. From the theoretical point of view it is interesting to
consider the problem for the case of the strong interaction when the motion
of a molecule in the angular space has the purely diffusive character. The
corresponding analysis will be published elsewhere.

7. So far we have considered the molecule to be embedded inside bulk helium.
In the case of \ a cluster of finite size L the results obtained for the
Fermi liquid require that the weak condition $k_{F}L>>1$ is fulfilled. This
is due to the fact that $k\thicksim k_{F}$ for the excitations
characteristic for the problem . In the case of a Bose liquid the
requirement is much stronger $\frac{\varepsilon _{01}}{\hbar u}L>>1.$ For
the opposite inequality there are no phonons in the system for which the
conservation energy law can be satisfied in Eq. (6). In this case the
interaction with the surface excitations, for which the wave function
penetrates effectively in the depth of \ a cluster, becomes significant.
This interesting problem will be considered separately.

\ \

\end{document}